\pdfoutput = 1
\documentclass[11pt]{article}
\usepackage{url}
\usepackage{graphicx}
\usepackage{amsfonts}
\usepackage{amssymb}
\usepackage{latexsym}
\usepackage{color}
\usepackage{multirow}
\usepackage[normalem]{ulem}

\newcommand{\notyet}[1]{}
\newcommand{\squeezelist}{\setlength{\itemsep}{0pt}}

\newcommand{\ABox}{
\raisebox{3pt}{\framebox[6pt]{\rule{6pt}{0pt}}}
}

\newtheorem{theorem}{{\bf Theorem}}

\newtheorem{property}[theorem]{Property}

\def\P{{\mathcal P}}
\def\c{\mathtt{C}}
\def\cc{\mathtt{CC}}

\newcommand\connect[1]{\texttt{connect}}
\newcommand\hand{{\sc Hand}}
\newcommand\unfold{{\sc UnfoldChild}}
\newcommand\unfoldret{{\sc UnfoldChildRetrace}}
\newcommand\unfoldrev{{\sc UnfoldChildReverse}}

\newcommand\unfoldrim{{\sc UnfoldRimChildren}}
\newcommand\unfoldrims{{\sc Unfold}}

\begin{document}

\title{Unfolding Orthogrids with Constant Refinement}
\author{
Mirela Damian\thanks{Villanova University, \tt{mirela.damian@villanova.edu}}
\and
Erik Demaine\thanks{Massachusetts Institute of Technology,  \tt{edemaine@mit.edu}}
\and
Robin Flatland\thanks{Siena College, \tt{flatland@siena.edu}}
}
\date{}
\maketitle

\begin{abstract}
We define a new class of orthogonal polyhedra, called \emph{orthogrids}, that can be unfolded without overlap with constant refinement of the gridded surface. 
\end{abstract}

\section{Introduction}
An \emph{unfolding} of a polyhedron is obtained by cutting the surface of the polyhedron and flattening it in the plane 
as a simple non-overlapping polygon. 
An \emph{edge unfolding} considers only cuts made along edges, while general unfoldings allow cuts anywhere on the surface. 

A polyhedron is \emph{orthogonal} if each of its faces is perpendicular to a coordinate axis.
An orthogonal polyhedron cannot be unfolded using edge cuts alone 
(the simplest example is a box sitting on top of a larger box). 
A \emph{grid unfolding} of a polyhedron $P$ allows additional cuts along grid edges created by slicing $P$ with 
axis-aligned planes that pass through at least one polyhedron vertex, which we refer to as \emph{grid planes}. Few nontrivial subclasses of orthogonal polyhedra are known to have grid unfoldings: orthotubes
\cite{Biedl-Demaine-Demaine-Lubiw-Overmars-O'Rourke-Robbins-Whitesides-1998},
orthostacks composed of orthogonally convex slabs
\cite{Damian-Meijer-2004-orthostacks}, 
well-separated orthotrees 
\cite{Damian-Flatland-Meijer-O'Rourke-2005-orthotrees} and
terrains
\cite{O'Rourke-2007-terrains}.

A $k$-\emph{refined grid unfolding}, defined for an orthogonal polyhedron and for some integer $k > 1$, allows additional cuts created by at most $k$ planes parallel to and sandwiched between adjacent grid planes. 
A remarkable breakthrough is the $k$-refined grid unfolding of any orthogonal polyhedron homeomorphic to a sphere, where $k$ is polynomial in the number of polyhedron vertices~\cite{Damian-Demaine-Flatland-2012-epsilon}.
For constant $k$, only two  classes of orthogonal polyhedra are known to have a $k$-refined grid unfolding: orthostacks
\cite{Biedl-Demaine-Demaine-Lubiw-Overmars-O'Rourke-Robbins-Whitesides-1998},
and Manhattan towers \cite{Damian-Flatland-O'Rourke-2008-manhattan}. 

Informally, an orthogonal polyhedron $P$ is an \emph{orthogrid} if it is composed of extrusions of simple polygons (which we call \emph{slabs}) stacked together so that a (left) vertex belongs to a unique slab. It turns out that this restriction forces adjacent slabs to cross each other and collectively form a (non-uniform) orthogonal grid structure, hence the name of ``orthogrid''. See ahead to Fig.~\ref{fig:defs2} for two orthogrid examples. 

\paragraph{Our Result.}  In this paper we introduce a new class of orthogonal polyhedra, called \emph{orthogrids}, and show that orthogrids have a $k$-refined grid unfolding for some constant integer $k > 1$. Our restriction to orthogrids is necessary to guarantee that some front/back surface piece is available at each vertex, so it can be used in unfolding as needed. 

\section{Definitions}
%\scalebox{1.5}{$\looparrowright$}indicator
%
Let $\P$ be an orthogonal polyhedron with the surface homeomorphic to a sphere (i.e., genus zero).
% $A$ stands for parent band later.
Classify the faces of $\P$ according to the direction of their outward normal:
\emph{front} ($+y$), \emph{back} ($-y$), \emph{top} ($+z$),
\emph{bottom} ($-z$), \emph{right} ($+x$) and \emph{left} ($-x$). 
The vertices of $\P$ fall into similar categories, with the difference 
that a top or bottom vertex is also a left or right vertex, depending on its two incident faces. 
We take the $z$-axis
to define the vertical direction; \emph{vertical} faces are parallel to the $xz$-plane
or the $yz$ plane.  Clockwise (cw) and counterclockwise (ccw) directions are defined with
respect to a viewpoint at $y = \infty$. 
The \emph{gridded} version of $\P$ adds edges to the surface of $\P$ by
intersecting $\P$ with planes parallel to Cartesian coordinate planes
through every vertex. We distinguish between an original vertex of the polyhedron,
which we call a \emph{corner vertex} or simply a \emph{vertex}, and the points of the grid (some of which may be vertices). 
%\emph{gridpoint}, a vertex of the grid (which might be an original vertex). A {\em gridedge} ({\em gridface}) is an edge (face) of the grid that lies on the surface of the polyhedron.

Let $Y_1, Y_2, \ldots$ be planes passing through every vertex of $\P$,
orthogonal to the $y$-axis. By convention, $Y_1$ is the plane with the
largest $y$-coordinate.
\emph{Layer} $i$ is the portion of $\P$ bounded by planes $Y_i$
and $Y_{i+1}$, and is composed of one or more connected
components called \emph{slabs}.
The \emph{front} (\emph{back}) face of a slab is the face with normal pointing
in the $+y$ ($-y$) direction, and it may be partly on the surface of $\P$ and partly
interior to $\P$. The top, right, bottom and left faces surrounding a slab form
a \emph{band}. The intersection between a band
$B$ and a bounding plane $Y_i$ is a cycle of edges called a \emph{rim},
which we denote by $r_i(B)$.
Each band $B$ in a layer $i$ has two rims, $r_i(B)$ and $r_{i+1}(B)$.
We use the notation 
%The \emph{rimfacet} 
$r_i[B]$ to denote the closed area enclosed by rim $r_i(B)$. 
(Note that $r_i[B]$ is either the front or the back face of the slab surrounded by $B$.)
Two bands $A$ and $B$ separated by a plane $Y_i$ are \emph{adjacent} if there is a point common to $r_i[A]$ and $r_i[B]$.

\begin{figure}[htbp]
\centering
\includegraphics[width=0.9\linewidth]{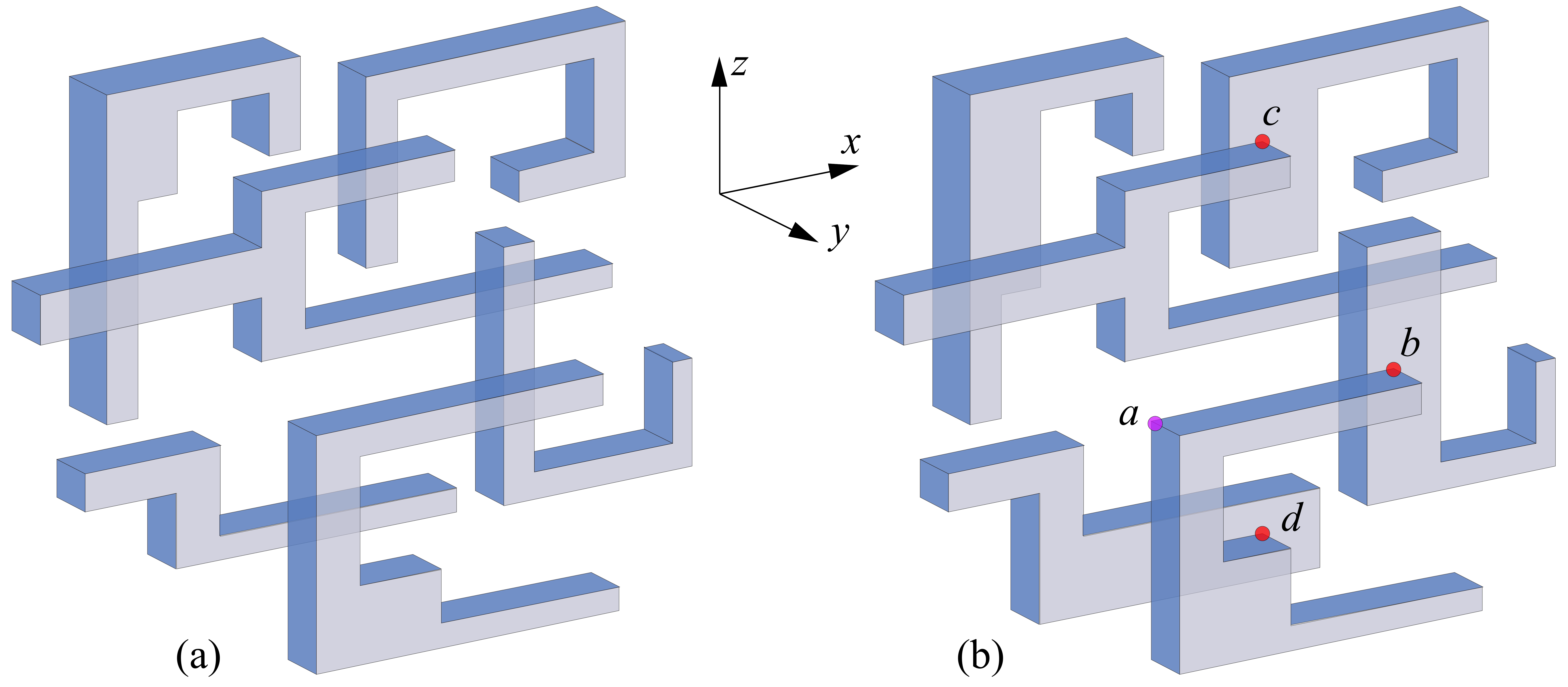}
\caption{Orthogrid with (a) all vertices exposed, (b) right vertices $b, c, d$ unexposed.}
\label{fig:defs2}
\end{figure}

If a vertex $u$ belongs to single slab of $\P$, then $u$ is called \emph{exposed}; otherwise, $u$ is \emph{unexposed.} In the example from Fig.~\ref{fig:defs2}b for instance, vertex $a$ is exposed, but vertices $b$, $c$ and $d$ are unexposed.
%
%%%%%%%%%%%%%%%%%%%%%%%%%%%%%%%%%Figure Begin
%%%%%%%%%%%%%%%%%%%%%%%%%%%%%%%%%Figure End
%

\paragraph{Orthogrids.} A genus zero orthogonal polyhedron $\P$ is an \emph{orthogrid} if all left vertices of $P$ are exposed. We further assume that $\P$ has no \emph{y-dents}, meaning that each $y$-perpendicular 2D cross section of $P$ has genus zero. 
Note that the meaning of ``left'' in the definition of an orthogrid is tied to a coordinate system,  therefore a polyhedron with all right vertices exposed  is also an orthogrid (simply reverse the direction of the $y$-axis). Fig.~\ref{fig:defs2} shows two examples of orthogrids: one with all vertices exposed (a), and one with some right vertices unexposed (b). 

%We show that any corner-disjoint orthocluster has a $k$-refined grid unfolding, for some constant $k > 1$. 

\section{Preliminary: Single Band Unfolding}
\label{sec:boxunf}
Single band unfolding is a basic tool used in several unfolding papers~\cite{Damian-Flatland-O'Rourke-2007-epsilon, Damian-Flatland-O'Rourke-2008-manhattan, Damian-Demaine-Flatland-2012-epsilon}. 
%For simplicity, we present the unfolding method for the case when the band is a rectangular box; the extension to bands of arbitrary shapes is immediate. 
The key idea is to identify a path on the box surface that cycles around the top, right, bottom and left faces of the box at least once, and can be flattened out into an $x$-monotone orthogonal path in the plane. We refer to this path as the \emph{unfolding spiral}, or simply the \emph{spiral}; we denote the spiral on the box surface $\xi_{3d}$, and the flattened version of the spiral in the plane $\xi_{2d}$.

%
%%%%%%%%%%%%%%%%%%%%%%%%%%%%%%%%%Figure Begin
\begin{figure}[htbp]
\centering
\includegraphics[width=0.9\linewidth]{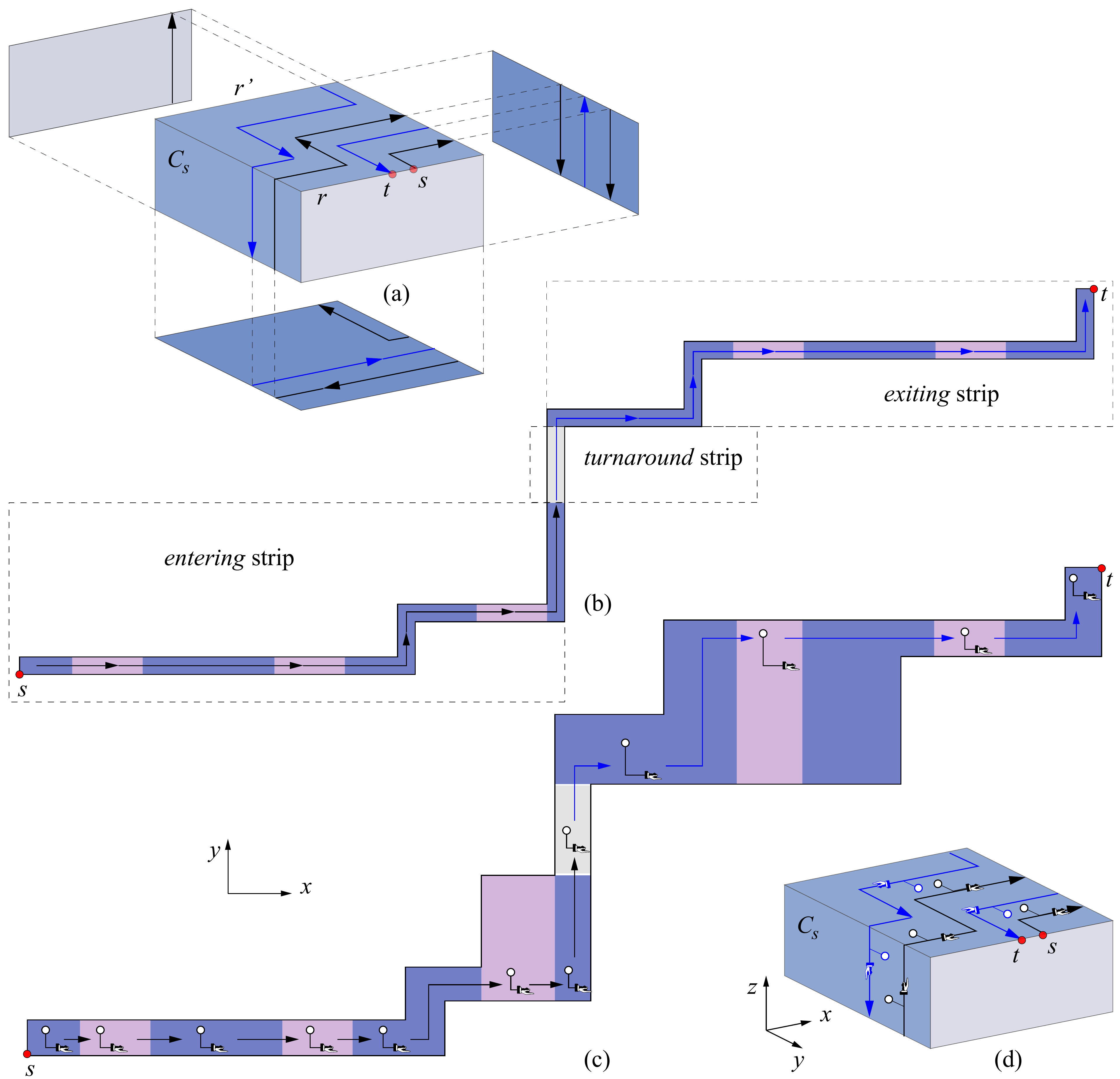}
\caption{Box unfolding (a) Unfolding spiral $\xi_{2d}$.
(b) Flattened spiral $\xi_{2d}$, slightly thickened to mark face transitions (c) Thickened spiral that covers the entire band (d) 3D \hand\ Orientation.}
\label{fig:Unf0}
\end{figure}
%%%%%%%%%%%%%%%%%%%%%%%%%%%%%%%%%Figure End
%

Figure~\ref{fig:Unf0}a shows the directed path followed by $\xi_{3d}$: it starts
at the entering point $s$ on the top edge of the front rim $r$ and spirals in clockwise
direction around the top, right, bottom, and left faces toward the back rim $r'$.
We call this spiral strip from $s$ up to the point it hits $r'$ the \emph{entering strip}.
We require that the entering strip cycles around the band \emph{at least once}, before
moving towards the back rim $r'$.
When it reaches $r'$, $\xi_{3d}$ turns around by crossing the back face upwards toward
the top edge of $r'$; we refer to this back face strip as the \emph{turnaround} strip, to emphasize its role in the unfolding. From $r'$, the unfolding path $\xi_{3d}$ spirals in a counterclockwise direction back to $r$, ending at
the exiting point $t$ lying next to $s$ on the front rim. We call this strip piece extending
from $r'$ back to $t$ the \emph{exiting strip}. Thus the entering strip and the exiting
strip spiral in opposite directions.

When $\xi_{3d}$ is cut out, unfolded and laid horizontally
in the plane, it forms a monotonic staircase $\xi_{2d}$ as depicted in Fig.~\ref{fig:Unf0}b. The spiral can be subsequently thickened so that it covers the top, right, bottom and front faces entirely, as depicted in Fig.~\ref{fig:Unf0}c. Marked along the flattened spiral path in Fig.~\ref{fig:Unf0}d is an L-shaped
indicator (or simply an $L$-indicator) with two pointers corresponding to the coordinate axes in the plane:
the \emph{hand pointer} pointing in the direction of the positive $x$-axis, and the
\emph{head pointer} pointing in the direction of the positive $y$-axis. When mapped back onto the
3D box shape, the $L$-indicator traces the path taken by $\xi_{3d}$. This makes it easier to visualize the orientation of $\xi_{2d}$ just by looking at the $L$-indicator on $\xi_{3d}$: if the $L$-indicator glides in the direction of the hand pointer along a piece of $\xi_{3d}$, that piece flattens out horizontally to the right in the plane; if the $L$-indicator glides in the direction opposite to the hand pointer along a piece of $\xi_{3d}$, that piece flattens out horizontally to the left in the plane. This is what we will try to avoid (with small exceptions) so that $\xi_{2d}$ maintains its $x$-monotonicity. If the $L$-indicator glides in a direction orthogonal to the hand pointer along a piece of $\xi_{3d}$, that piece flattens out vertically in the plane. In this latter case, the up or down direction is irrelevant, because we seek monotonicity only in the horizontal direction. For this reason, we will henceforth refer to the $L$-indicator as \hand, to remind the reader that it is really the hand pointer that affects the monotonicity property of $\xi_{2d}$. Nevertheless, the head pointer will be useful in tracing the paths taken by the unfolding spiral on the surface of $\P$. In Fig.~\ref{fig:Unf0}d for example, one can imagine the \hand\ sliding across the surface in the direction indicated by the arrows; the head pointer is simply an aid in this visualization process.

%
%%%%%%%%%%%%%%%%%%%%%%%%%%%%%%%%%Figure Begin
\begin{figure}[htbp]
\centering
\includegraphics[width=0.9\linewidth]{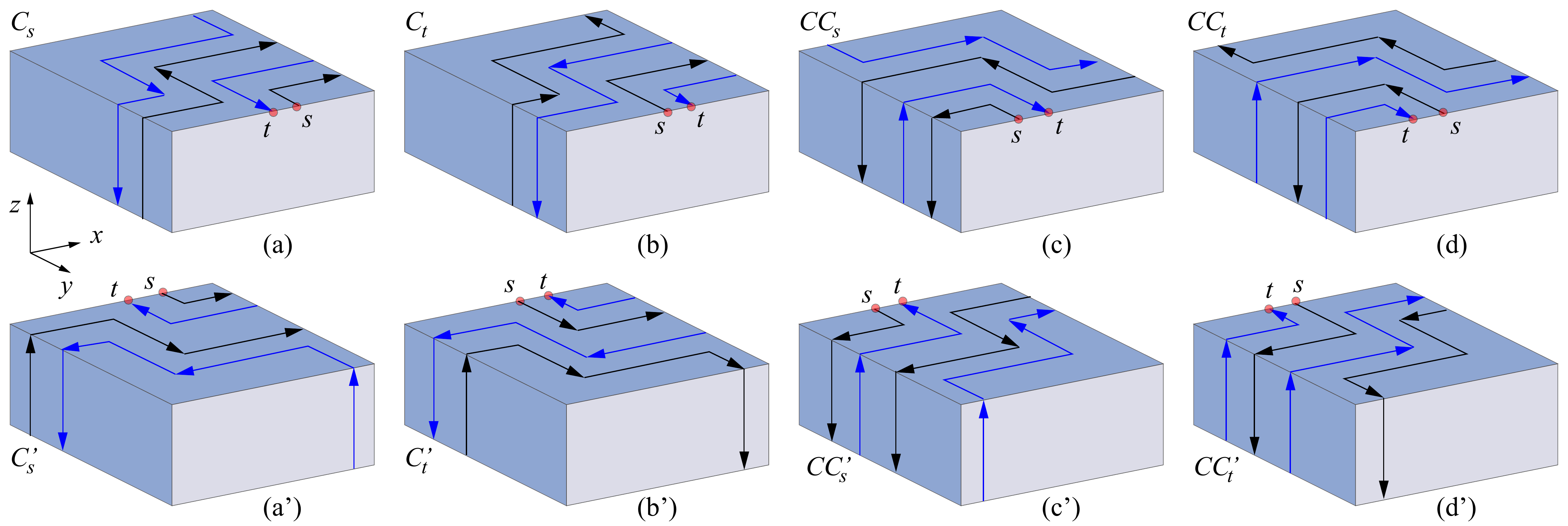}
\caption{Types of unfolding (a) $\c_{s}$ (b) $\c_{t}$ (c) $\cc_{s}$ (d) $\cc_{t}$ (a$'$) $\c'_{s}$ (b$'$) $\c'_{t}$ (c$'$) $\cc'_{s}$ (d$'$) $\cc'_{t}$.}
\label{fig:Unf}
\end{figure}
%%%%%%%%%%%%%%%%%%%%%%%%%%%%%%%%%Figure End
%

The type of unfolding depicted in Fig.~\ref{fig:Unf0} and again in Fig.~\ref{fig:Unf}a is denoted by $\c_s$. The symbol $\c$ is used to indicate that the entering strip starts on the front rim of the band and cycles clockwise around the band. The subscript $\_s$ is used to indicate that the (black)
entering strip (rather than the (blue) exiting strip) runs alongside the rim containing $s$ and $t$; the only place it doesn't run alongside the rim is at the small gap at $t$ where the exiting spiral leaves the band.
%begins by making an ``almost complete'' cycle around the rim containing
%$s$, leaving a small gap for the exiting strip.
Sometimes the unfolding of a band $A$ will be required to begin on the back rim of $A$; we use the symbol $\c'$ to indicate that the unfolding begins on the back rim. The $\c'_s$ unfolding is simply an $xz$-reflection of the $\c_s$ unfolding, and is depicted in Fig.~\ref{fig:Unf}a$'$. The only difference between $\c_s$ and $\c'_s$ is that $s$ and $t$ are located on the front rim in the first case, and on the back rim in the latter case; the fact that the entering strip starts by making a clockwise cycle around the rim containing $s$ applies to both cases.

Alternate types of unfolding are $\c_t$ (Fig.~\ref{fig:Unf}b), $\cc_s$ (Fig.~\ref{fig:Unf}c), $\cc_t$ (Fig.~\ref{fig:Unf}d), and their corresponding primed $xz$-reflections $\c'_t$ (Fig.~\ref{fig:Unf}b$'$), $\cc'_s$ (Fig.~\ref{fig:Unf}c$'$) and $\cc'_t$ (Fig.~\ref{fig:Unf}d$'$). Here the symbol $\cc$ ($\cc'$) is used to indicate that the entering strip starts on the front (back) rim of the band and cycles counter-clockwise around the band; and the subscript $\_t$ is used to indicate that
the (blue) exiting strip runs alongside the rim containing $s$ and $t$ (with the exception of the small gap between $s$ and $t$).
%
%the entering strip does not cycle along the rim $r$ containing $s$, but instead
%it takes a step towards the other rim of the band, leaving room for the exiting
%trip to make a complete cycle around $r$.

For ease of reference we introduce the following definitions. The band rim containing the entering point $s$ is called the \emph{entering rim}; the opposite band rim is called the \emph{turnaround} rim, to indicate the fact that the turnaround is executed on the facet of that rim. Note that, if the unfolding type for a band is $\c$ or $\cc$, the entering rim is the front rim and the turnaround rim is the back rim; and if the unfolding type is $\c'$ or $\cc'$, then the entering rim is the back rim and the turnaround rim is the front rim. 

Call the strip that runs alongside the front rim of a band the \emph{front strip}. Note that the front strip could be either on the entering spiral
%(case of $\_s$ unfolding)
or on the exiting spiral.
%(case of $\_t$ unfolding).
Similarly, call the strip that runs alongside the back rim of a band the \emph{back strip}. The unfolding directions of the front and back strips corresponding to the eight unfolding types depicted in Fig.~\ref{fig:Unf} are summarized in Table~\ref{tab:FrontBack}.
\begin{table}[htpb]
\centering{
\begin{tabular}{|c|c|c|}
\hline
Unfolding Label & Front Strip & Back strip \\
\hline\hline
$\c_{s}$ & Clockwise, on entering spiral & Counterclockwise, on exiting spiral\\
\hline
$\c'_{s}$ & Counterclockwise, on exiting spiral & Clockwise, on entering spiral\\
\hline
$\c_{t}$ & Counterclockwise, on exiting spiral & Clockwise, on entering spiral\\
\hline
$\c'_{t}$ & Clockwise, on entering spiral & Counterclockwise, on exiting spiral\\
\hline
$\cc_{s}$ & Counterclockwise, on entering spiral & Clockwise, on exiting spiral\\
\hline
$\cc'_{s}$ & Clockwise, on exiting spiral & Counterclockwise, on entering spiral\\
\hline
$\cc_{t}$ & Clockwise, on exiting spiral & Counterclockwise, on entering spiral\\
\hline
$\cc'_{t}$ & Counterclockwise, on entering spiral & Clockwise, on exiting spiral\\
\hline
\end{tabular}
\caption{Unfolding directions for front and back strips.}
\label{tab:FrontBack}
}%centering
\end{table}

\section{Overview of Unfolding Algorithm}
\label{sec:unfidea}
Observe that in the case of an orthogrid, any two adjacent bands exhibit a left face orthogonally crossing a top face of the other band. (This is because all left vertices of an orthogrid are exposed.) Our unfolding algorithm uses this property in transitioning from one band to the other and back. 
%It turns out that all algorithmic issues present in the unfolding algorithm also arise in the case where each slab is just a rectangular box (see the example depicted in Fig.~\ref{fig:defs}). Therefore we will describe the unfolding algorithm for this simple shape class first, and then show that the ideas extend directly to unfolding all orthogrids.
%
%%%%%%%%%%%%%%%%%%%%%%%%%%%%%%%%%Figure Begin
%\begin{figure}[htbp]
%\centering
%\includegraphics[width=0.45\linewidth]{defs}
%\caption{An orthogrid with all slabs shaped as rectangular boxes.}
%\label{fig:defs}
%\end{figure}
%%%%%%%%%%%%%%%%%%%%%%%%%%%%%%%%%Figure End
%
The overall structure of the unfolding algorithm is as follows:
\begin{enumerate}
\item Compute a band unfolding tree $T$. Select the root of $T$ to be a band of smallest y-coordinate, with ties broken arbitrarily (\S\ref{sec:unftree}).
\item Determine forward and return transition paths corresponding to each arc in $T$, and assign an unfolding label to each band in $T$ (\S\ref{sec:transitions}).
\item Starting with the root band, unfold each band as a conceptual unit, but interrupt the unfolding each time a forward transition path to a child is encountered. The unfolding follows the forward transition path to the child band, the child band is recursively unfolded, then the unfolding returns along the return transition  path back to the parent, resuming the parent band unfolding from the point it left off. The resulting strip is laid out horizontally in the plane (\S\ref{sec:alg}).
\item Attach the vertical front and back faces of $\P$ below and above appropriate horizontal sections of the unfolding strip determined in the previous step (\S\ref{sec:frontback}).
\end{enumerate}
We proceed to presenting the details of each of these steps.

\section{Computing an Unfolding Tree $T$}
\label{sec:unftree}
%Because all left vertices of an orthogrid are exposed, two adjacent bands of $\P$ always involve a left face of one band orthogonally crossing a top face of the other band. 
The unfolding tree $T$ has one node for each band and anchors connecting pairs of adjacent bands as follows. For each band $A$ and each band $B$ adjacent to $A$, we add arc $(A,B)$ to $T$ and select the \emph{anchor} point connecting $A$ to $B$ to be the  leftmost topmost intersection point between $A$ and $B$.  %Note that such a point always exists at the intersection between a left face of one band and a top face of the other band (because all left vertices of an orthogrid are exposed).  
It can be verified that the structure $T$ constructed this way os a tree that spans all bands.

%\begin{lemma}
%$T$ is a tree that spans all bands.
%\label{lem:unftree}
%\end{lemma}
%
To distinguish edges in $T$ from edges of the orthogonal polyhedron $\P$, we will refer to
the first as \emph{arcs}, and to the latter simply as \emph{edges}. A child is a \emph{front child} (\emph{back child}) if it is adjacent to the front (back) of its parent.

%
%%%%%%%%%%%%%%%%%%%%%%%%%%%%%%%%%Figure Begin
\begin{figure}[htbp]
\centering
\includegraphics[width=0.75\linewidth]{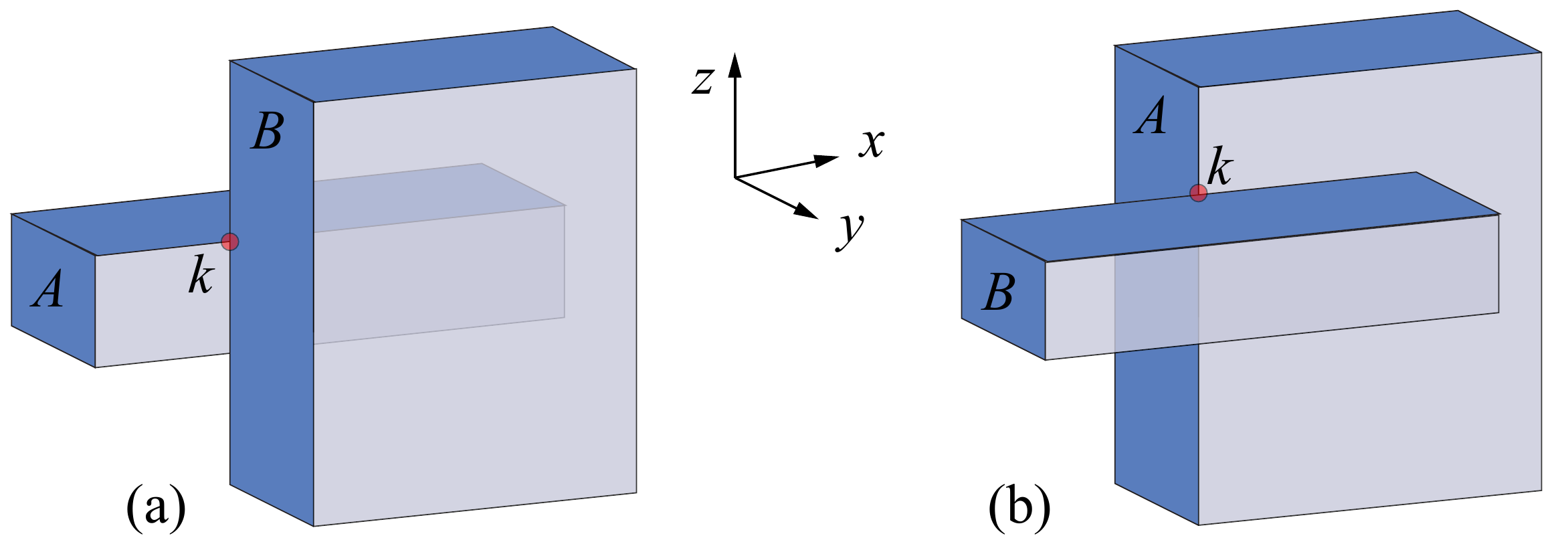}
\caption{Anchor classification (a) Class C1 (b) Class C2.}
\label{fig:anchors}
\end{figure}
%%%%%%%%%%%%%%%%%%%%%%%%%%%%%%%%%Figure End

\noindent
An anchor connecting a parent band $A$ to a child band $B$ falls in one of the following two categories:
\begin{itemize}
\item [(C1)] The anchor is a point $k$ at the intersection between a horizontal edge of $A$ and a vertical edge of $B$ 
(see Fig.~\ref{fig:anchors}a). We call $k$ a C1-\emph{anchor}, and $B$ a C1-\emph{child}. 
\item [(C2)] The anchor is a point $k$ at the intersection between a vertical edge of $A$ and a horizontal edge of $B$
(see Fig.~\ref{fig:anchors}b). We call $k$ a C2-\emph{anchor}, and $B$ a C2-\emph{child}. 
\end{itemize}

\section{Unfolding Tools}
In this section we present three key tools used in implementing the recursive unfolding procedure sketched in \S\ref{sec:unfidea}: \emph{forced turns} (\S\ref{sec:forcedturns}), \emph{unfolding variations} (\S\ref{sec:kinunf}) and \emph{forward and return transition paths} (\S\ref{sec:transitions}).
In the unfolding procedure of $\P$, we say that a band $B$ has been \emph{visited} if an unfolding path has been traced on $B$ by \hand. We seek to maintain two invariants throughout the unfolding procedure:
\begin{enumerate}
\item [(I1)] The \hand\ is parallel to the $x$-axis on a horizontal face of a band, pointing in the direction of the unfolding.
\item [(I2)] If a vertical rim edge $e$ of a band $A$ contains at least one anchor point attaching an \emph{unvisited} child of $A$ to $A$, then the \hand\ is parallel to the $y$-axis (pointing to either $+y$ or $-y$) while moving along $e$.
\end{enumerate}
As we shall shortly see, invariant (I2) enables recursive unfolding of $A$'s children attached to $e$.

\subsection{Forced Turns}
\label{sec:forcedturns}
Invariants (I1) and (I2) will sometimes require the \hand\ to change orientation when moving from a horizontal face to a vertical face of a band, or vice-versa. Such a reorientation corresponds to a $\pm 90^\circ$-turn of the strip $\xi_{2d}$ in the plane. (Recall that $\xi_{2d}$ is the flattened plane version of the unfolding strip $\xi_{3d}$.) We call such reorientations \emph{forced turns}, to distinguish them from the normal turns resulting from the basic unfolding method described in \S\ref{sec:boxunf}.

%%%%%%%%%%%%%%%%%%%%%%%%%%%%%%%%%Figure Begin
%
\begin{figure}[htbp]
\centering
\includegraphics[width=0.8\linewidth]{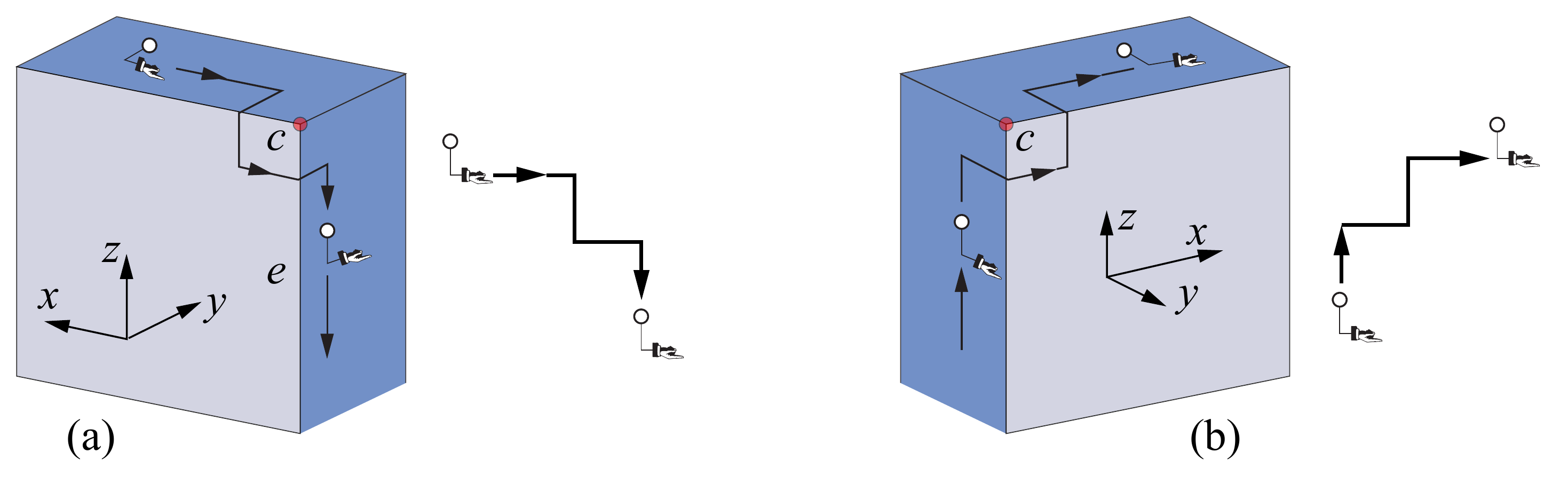}
\caption{Forced turns at convex vertex $c$ (a) horizontal-to-vertical, and  (b) vertical-to-horizontal.}
\label{fig:turns}
\end{figure}
%%%%%%%%%%%%%%%%%%%%%%%%%%%%%%%%%Figure End
%

Forced turns will be executed only at left vertices of $\P$, which by definition are exposed. Let $c$ be a convex vertex of a band $A$ where a forced turn occurs. The turn for the case when the \hand\ is on a horizontal (vertical) face of $A$ upon meeting $c$ is depicted in Fig.~\ref{fig:turns}a(b), for both $\xi_{3d}$ (left) and $\xi_{2d}$ (right). The coordinate axes from Fig.~\ref{fig:turns}a indicate that the top left vertex $c$ is encountered in a counterclockwise traversal of the band; however, a simple reorientation of the coordinate axes shows that this turn also applies to the case when $c$ is a  bottom left vertex encountered in a clockwise band traversal. 

%%%%%%%%%%%%%%%%%%%%%%%%%%%%%%%%%Figure Begin
%
\begin{figure}[htbp]
\centering
\includegraphics[width=0.8\linewidth]{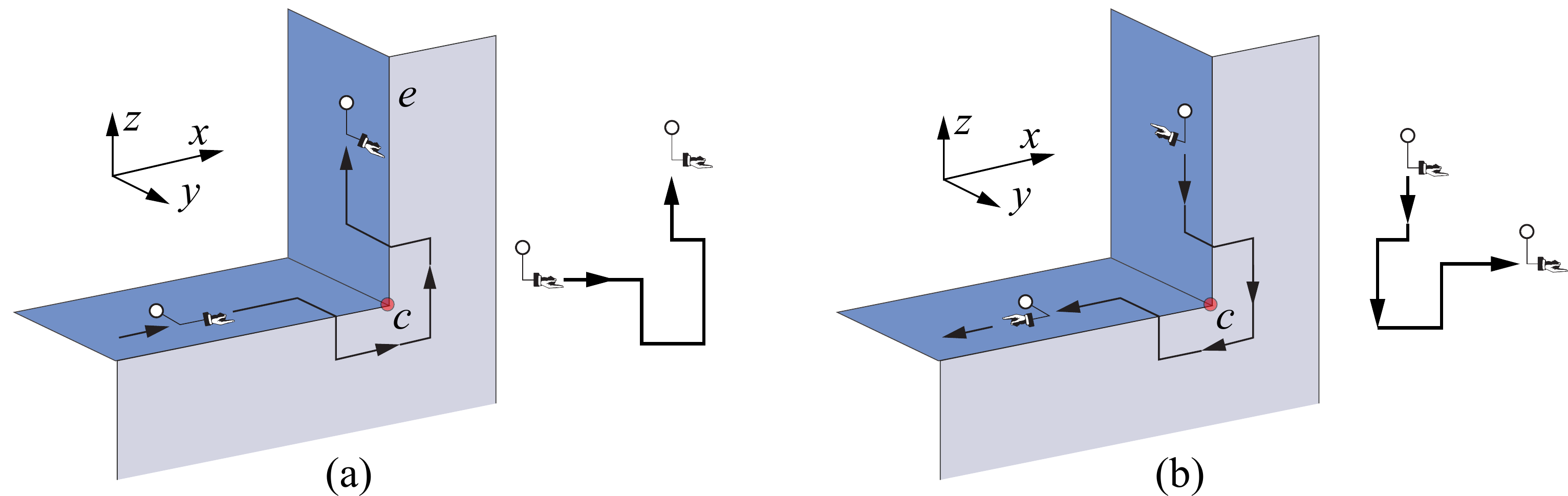}
\caption{Forced turns at reflex vertex $c$ (a) horizontal-to-vertical, and  (b) vertical-to-horizontal.}
\label{fig:turns2}
\end{figure}
%%%%%%%%%%%%%%%%%%%%%%%%%%%%%%%%%Figure End
%

Fig.~\ref{fig:turns2} a(b) shows the forced turn for the case where $c$ is a reflex vertex, and the \hand\ is on a horizontal (vertical) face of $A$ upon meeting $c$. Note that the paths traced by $\xi_{2d}$ in Fig.~\ref{fig:turns2} are not monotone in the $x$-direction, but the extent of the turn steps are adjusted so that $\xi_{2d}$ does not self-intersect (the steps can be arbitrarily small, but are exaggerated in the figure for the sake of clarity).  The forced turns depicted in Figs.~\ref{fig:turns}a and~\ref{fig:turns2}a indicate that the following property holds:
\begin{property}
Just after a forced horizontal-to-vertical turn, the \hand\ points away from the $Y$-plane containing $c$ if $c$ is convex (Fig.~\ref{fig:turns}a), and toward the $Y$-plane containing $c$ if $c$ is reflex (Fig.~\ref{fig:turns2}a).
\end{property}
Also observe that the clockwise/counterclockwise unfolding direction is the same before and after a forced turn. This is necessary so that the unfolding advances along the path depicted for the single band in \S\ref{sec:boxunf}, rather than backtracking. To make it possible for vertical-to-horizontal turns to also preserve the unfolding direction, we require one more invariant to be maintained throughout the unfolding procedure:
\begin{enumerate}
\squeezelist
\item[(I3)] Each vertical-to-horizontal forced turn at a corner vertex $c$ reorients the \hand\ parallel to the $x$-axis, pointing in the the unfolding direction. Just before a vertical-to-horizontal turn is forced at a vertex $c$, the \hand\ points toward the $Y$-plane containing $c$ if $c$ is convex (Fig.~\ref{fig:turns}b), and away from the $Y$-plane containing $c$ if $c$ is reflex (Fig.~\ref{fig:turns2}b).
\end{enumerate}
Note that the vertical-to-horizontal turns depicted in Figs.~\ref{fig:turns}b and~\ref{fig:turns2}b for $\xi_{2d}$ are mirrors of the corresponding horizontal-to-vertical turns from Figs.~\ref{fig:turns}a and~\ref{fig:turns2}a, with the unfolding direction reversed. 
Also note that each horizontal-to-vertical forced turn results in the \hand\ pointing in a direction opposite to the one required to 
perform the counterpart vertical-to-horizontal turn (compare Figs.~\ref{fig:turns}a and~\ref{fig:turns}b, and similarly for 
Figs.~\ref{fig:turns2}a and~\ref{fig:turns2}b). As we shall later see,
a horizontal-to-vertical turn at a corner vertex $c$ of a band $A$ is performed only if an unvisited
child of $A$ is attached to the vertical edge $e$ of $A$ incident to $c$ (so a forced turn is necessary to maintain invariant I2). This will enable us to use one such unvisited child to reverse the direction of the \hand\, if necessary. 

Throughout the paper, the terms \emph{reverse} and \emph{opposite} associated with a vector $\overrightarrow{u}$ indicate the scalar $-1$ multiplied by $\overrightarrow{u}$. 

%\subsection{Kindred Unfoldings}
\subsection{Unfolding Variations}
\label{sec:kinunf}
In this section we present one variation to the basic unfolding types presented in \S\ref{sec:boxunf}, to accommodate for the fact that sometimes the unfolding of a band must begin on a vertical face of the band. We refer to the unfolding types depicted in Fig.~\ref{fig:Unf} as $x$-unfoldings, to indicate the fact that the unfolding starts and ends with the \hand\ parallel to the $x$-axis. We introduce a variation called $y$-\emph{unfolding}, which begins and ends with the \hand\ parallel to the $y$-axis (see ahead the entering and exiting hand in $B$'s unfolding from Fig.~\ref{fig:C1-transitions}). 
%%%%%%%%%%%%%%%%%%%%%%%%%%%%%%%%%Figure Begin
\begin{figure}[htbp]
\centering
\includegraphics[width=0.25\linewidth]{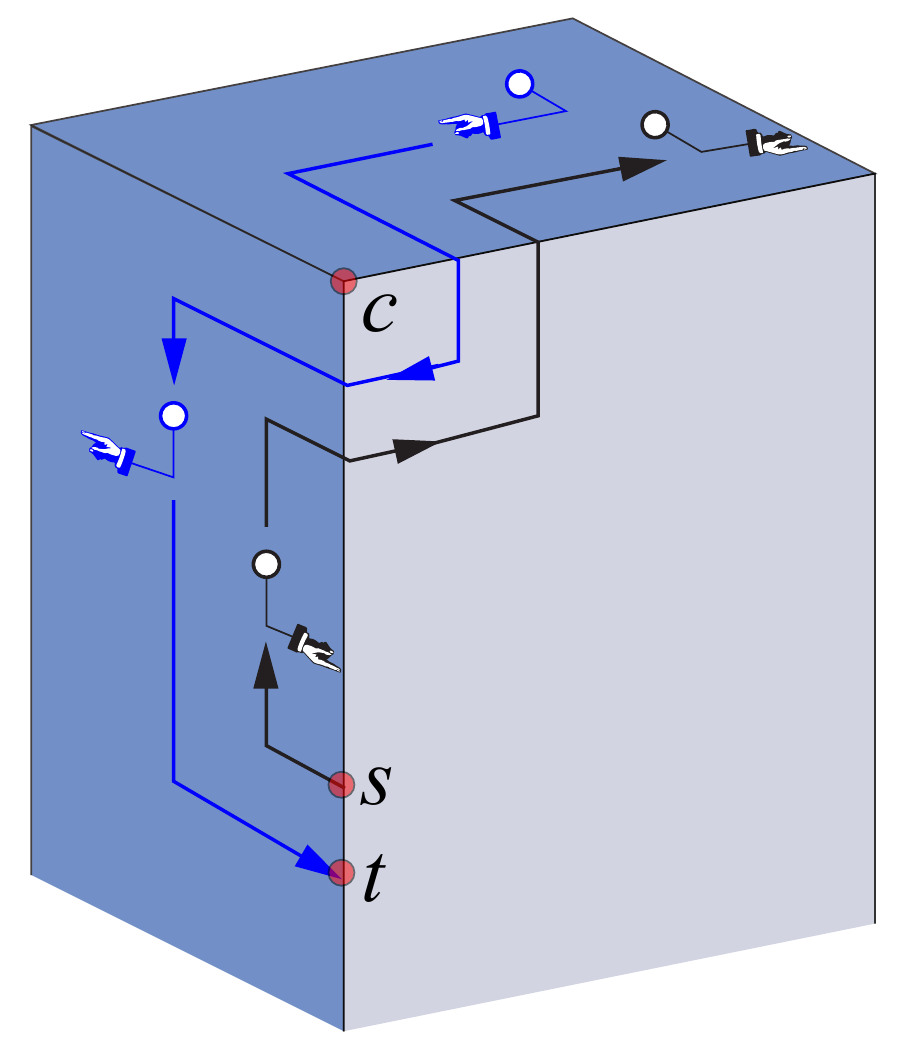}
\caption{Retracing forced turns.}
\label{fig:reverseturns}
\end{figure}
%%%%%%%%%%%%%%%%%%%%%%%%%%%%%%%%%Figure End

Consider an unfolding that starts on a vertical face of a band, with the \hand\ parallel to the $y$-axis (see starting point marked $s$ in Fig.~\ref{fig:reverseturns}). 
Let $c$ be the first vertex encountered by the \hand\ during the unfolding. Our discussion for the rest of this section is based on the assumption that invariant (I3) holds, meaning that upon reaching $c$, the \hand\ is oriented in such a way that a forced turn at $c$ maintains the unfolding direction. After the forced turn, the unfolding spiral follows the same path as in the case of an $x$-unfolding, up to the point where the exiting strip reaches vertex $c$ on its return trip. At this point the exiting strip retraces the path taken by the entering strip around $c$ and along the vertical face incident to $c$.% so as to reorient the exiting \hand\ to be opposite to the entering \hand. 

Fig.~\ref{fig:reverseturns} shows the retracing path around a convex vertex $c$ (the reflex vertex case is similar). 
Note that the nesting of the unfolding strips around $c$ matches the nesting of the strips on the incident band faces, so no crossing occurs. Also note that turns taken in opposite directions at $c$ also have opposite \hand\ orientations on both incident (vertical and horizontal) band faces.

We remind the reader that the extent of the turn steps are exaggerated for clarity in our figures, but in fact the turns can be made arbitrarily close to the corner vertex. Similarly, the entering and exiting point can be placed arbitrarily close to each other. The following property will be crucial in proving the correctness of the unfolding algorithm.
\begin{property}
The entering \hand\ and the exiting \hand\ on a band have opposite orientations. 
\label{prop:handexit}
\end{property}

\subsection{Transition Paths and Unfolding Labels}
\label{sec:transitions}
Corresponding to each arc in $T$ connecting a parent band $A$ to a child band $B$, we select two transition paths: a \emph{forward transition path} from $A$ to $B$, and a \emph{return transition path} from $B$ back to $A$. When the unfolding strip for $A$ reaches a forward transition path to an unvisited child $B$, the unfolding follows this path to $B$, band $B$ is recursively unfolded, then unfolding follows the return transition path back to $A$. Each of these two transitions must maintain invariants (I1)-(I3). In addition, the return transition must be executed in such a way that the unfolding of $A$ can be resumed at the point it left off, as if the interruption did not occur. In other words, the orientation of the \hand\ on $A$ following the return transition must be identical to the one on $A$ prior to the forward transition.
This section is concerned with determining the forward and return connecting paths corresponding to each arc in $T$, and assigning an unfolding label to each band in $T$ that is consistent with these transitions.

Consider an arbitrary arc $(A, B) \in T$ connecting parent band $A$ to child band $B$. Let $Y_i$ be
the plane separating $A$ and $B$. In the following we discuss the transitions between $A$ and $B$ corresponding 
to each of the two anchor classes.

\subsubsection{C1-Transitions}
\label{sec:C1}
In this case the anchor connecting child $B$ to parent $A$ is a point $k$ at the intersection between a horizontal edge of $A$ and a vertical edge of $B$, as depicted in Fig.~\ref{fig:anchors}a. Consider first the case in which the unfolding proceeds clockwise along the top edge of $A$ upon reaching $k$.
By invariant (I1), the \hand\ is parallel to the $x$-axis on the top edge of $A$. Let $e$ be the vertical edge of $B$ incident to $k$. After the forward transition to $B$, the \hand\ must be parallel to the $y$-axis on $B$ (so that invariant (I2) is satisfied in case unvisited children of $B$ are attached along $e$) and pointing in the direction required by invariant (I3).

%
%%%%%%%%%%%%%%%%%%%%%%%%%%%%%%%%%Figure Begin
\begin{figure}[htpb]
\centering
\includegraphics[width=0.8\linewidth]{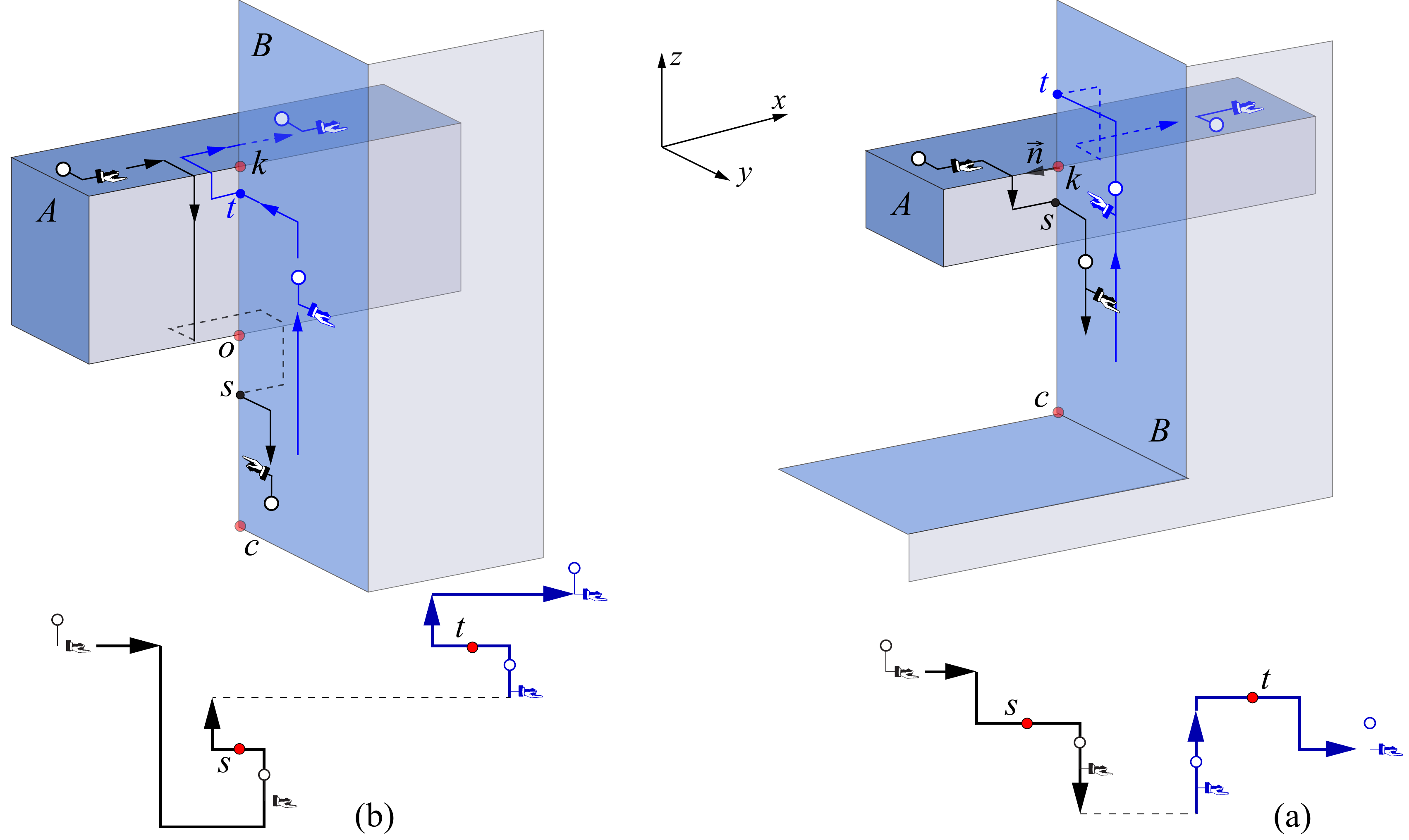}
\caption{C1-Transition paths (a) corner vertex $c$ unexposed (b) $c$ exposed.}
\label{fig:C1-transitions}
\end{figure}
%%%%%%%%%%%%%%%%%%%%%%%%%%%%%%%%%Figure End
%

The forward transitions from $A$ to $B$ are depicted in Fig.~\ref{fig:C1-transitions} (follow the paths from $A$ to $B$ in the direction indicated by the arrows, ignoring for the moment the return transition paths from $B$ to $A$). The forward transition depends on the nature of the bottom endpoint $c$ of $e$:
\begin{enumerate}
\item If $c$ is convex, then  the \hand\ must point toward $Y_i$ upon reaching $c$ (invariant I3). This is accomplished by the forward transition depicted in Fig.~\ref{fig:C1-transitions}a. (Flattened spiral $\xi_{2d}$ is shown below $\xi_{3d}$.) Note that the transition uses a second point $o$ at the at the intersection between $A$ and $B$ on the open vertical segment $\overline{kc}$, to flip the orientation of the \hand. The intersection point $o$ always exists, because the left vertex $c$ is exposed (by the definition of an orthogrid). 
\item If $c$ is reflex, then the \hand\ must point away from $Y_i$ upon reaching $c$ (invariant I3); this property is enforced along the entire edge segment $kc$ by the forward transition depicted in Fig.~\ref{fig:C1-transitions}b ($\xi_{2d}$ shown below $\xi_{3d}$).
\end{enumerate}
%the transition from $A$ and $B$ is the only unfolding event $o$ is involved in, so there are no conflicts over using $o$ in other events (forced turns or other transitions).}
%
For the cases depicted in Fig.~\ref{fig:C1-transitions}, 
the entry and exit points for $B$ are selected so that they support a recursive $y$-unfolding of $B$; these points are marked in Fig.~\ref{fig:C1-transitions} for each situation identified above. Note that upon exiting $B$ at $t$,  the \hand\ is parallel to the $y$-axis and pointing in the direction opposite to the one used upon entering $B$ at $s$ (by Property~\ref{prop:handexit}). This enables the unfolding strip to make the complementary return transition from $B$ back to $A$ depicted in Fig.~\ref{fig:C1-transitions} (follow the return path from $B$ to $A$ in the direction of the arrows). Note that each of these return transitions preserve the orientation of the \hand\  on $A$ as required by invariant (I3). 

Our discussion so far was based on the assumption that the unfolding proceeded clockwise along the top edge of $A$ before reaching $k$. This assumption helps visualize the transitions between the $A$ and $B$, however the transitions are general: the case when the unfolding strip proceeds counterclockwise on $A$ upon meeting $k$ is similar to the clockwise case, with forward and return path switching roles and traversed by the \hand\ in opposite direction. 

Next we assign an unfolding label to $B$ consistent with the transitions discussed above. The label we assign to $B$ depends 
%on the convexity of the vertex $c$, and 
on whether $B$ is a front or back child of $A$. 
We discuss three related cases; the unfolding labels for the second and third case will be derived from the unfolding label established in the first case.

\begin{table}[htpb]
\begin{small}
\centering{
\begin{tabular}{|c|c|c|c|}
\hline
Anchor Class & Child $B$ & Unfolding label for $A$ &  Unfolding label for $B$ \\
\hline\hline
\multirow{8}{*}{C1} &
\multirow{4}{*}{front} &
$\c_{s}$, $\c'_{t}$, $\cc'_{s}$, or $\cc_{t}$
&
\multirow{2}{*}{$\cc'_{s}$ (Figs.~\ref{fig:C1-transitions}a,b)} \\
&&(front strip clockwise)  & \\
\cline{3-4}
&& $\c'_{s}$, $\c_{t}$, $\cc_{s}$, or $\cc'_{t}$ & 
\multirow{2}{*}{$\cc'_t$} \\
&&(front strip counterclockwise) & \\
\cline{3-4}
\cline{2-4}
&\multirow{4}{*}{back} &
$\c'_{s}$, $\c_{t}$, $\cc_{s}$, or $\cc'_{t}$ & 
\multirow{2}{*}{$\cc_s$} \\
&&(back strip clockwise) & \\
\cline{3-4}
&& $\c_{s}$, $\c'_{t}$, $\cc'_{s}$, or $\cc_{t}$ & 
\multirow{2}{*}{$\cc_t$} \\
&&(back strip counterclockwise) & \\
\hline
\end{tabular}
\caption{Unfolding labels for $B$ when attached to $A$ by a C1-anchor. }
\label{tab:C1labels}
}%centering
\end{small}
\end{table}

\paragraph{Case 1.} $B$ is a front child of $A$ and the front strip of $A$ runs clockwise just before reaching $k$. We assign to $B$ the label $\cc'_{s}$ to indicate the fact that $B$'s unfolding starts at the back of $B$, and the entering strip runs alongside the back rim of $B$ in counterclockwise direction. Note that this label is consistent with the transitions from Fig.~\ref{fig:C1-transitions}.

\paragraph{Case 2.} $B$ is a front child of $A$ and the front strip of $A$ runs counterclockwise just before reaching $k$. Then the same transitions as in Case 1 occur, with the forward and return transition strips switching roles and traversed by the \hand\ in opposite direction. Note that the direction opposite to the one indicated by the return strip matches the direction of the forward strip, therefore the unfolding direction is as established for the clockwise case (Case 1). The only change in the unfolding label from Case 1 is switching indices $s$ and $t$, to indicate a switch in the roles of the entering and exiting strips. %(See Table~\ref{tab:C1labels}.)
Thus the unfolding label for $B$ is $\cc'_{t}$. 

\paragraph{Case 3.} $B$ is a back child of $A$. Note that an $xz$-mirror of Cases 1 and 2 above depicts the case in which $B$ is a back child of $A$; this transformation preserves the unfolding direction and the notion of left/right, but reverses the notion of front/back. 
Consequently, $B$'s unfolding label for the case when the back strip runs clockwise (counterclockwise) just before reaching $k$ can be obtained from $B$'s unfolding label for the case when the front rim runs clockwise (counterclockwise) just before reaching $k$, by replacing $\cc'$ by $\cc$. See Table~\ref{tab:C1labels} for a summary of all unfolding labels.

\subsubsection{C2-Transitions}
\label{sec:C2}
C2-transitions apply when the anchor connecting an unvisited child $B$ of parent $A$ is a point $k$ at the intersection between a vertical edge of $A$ and a horizontal edge of $B$, as depicted in Fig.~\ref{fig:anchors}b.
%Consider first the case in which $e_a$ is a left edge, and the unfolding proceeds clockwise along $e_a$.
By invariant (I2), the \hand\ is parallel to the $y$-axis upon reaching $k$. After making the forward transition to the horizontal face of $B$, the \hand\ must be parallel to the $x$-axis, so that invariant (I1) is satisfied. We now show how to execute such a transition.
%
%%%%%%%%%%%%%%%%%%%%%%%%%%%%%%%%%Figure Begin
\begin{figure}[hp]
\centering
\includegraphics[width=0.9\linewidth]{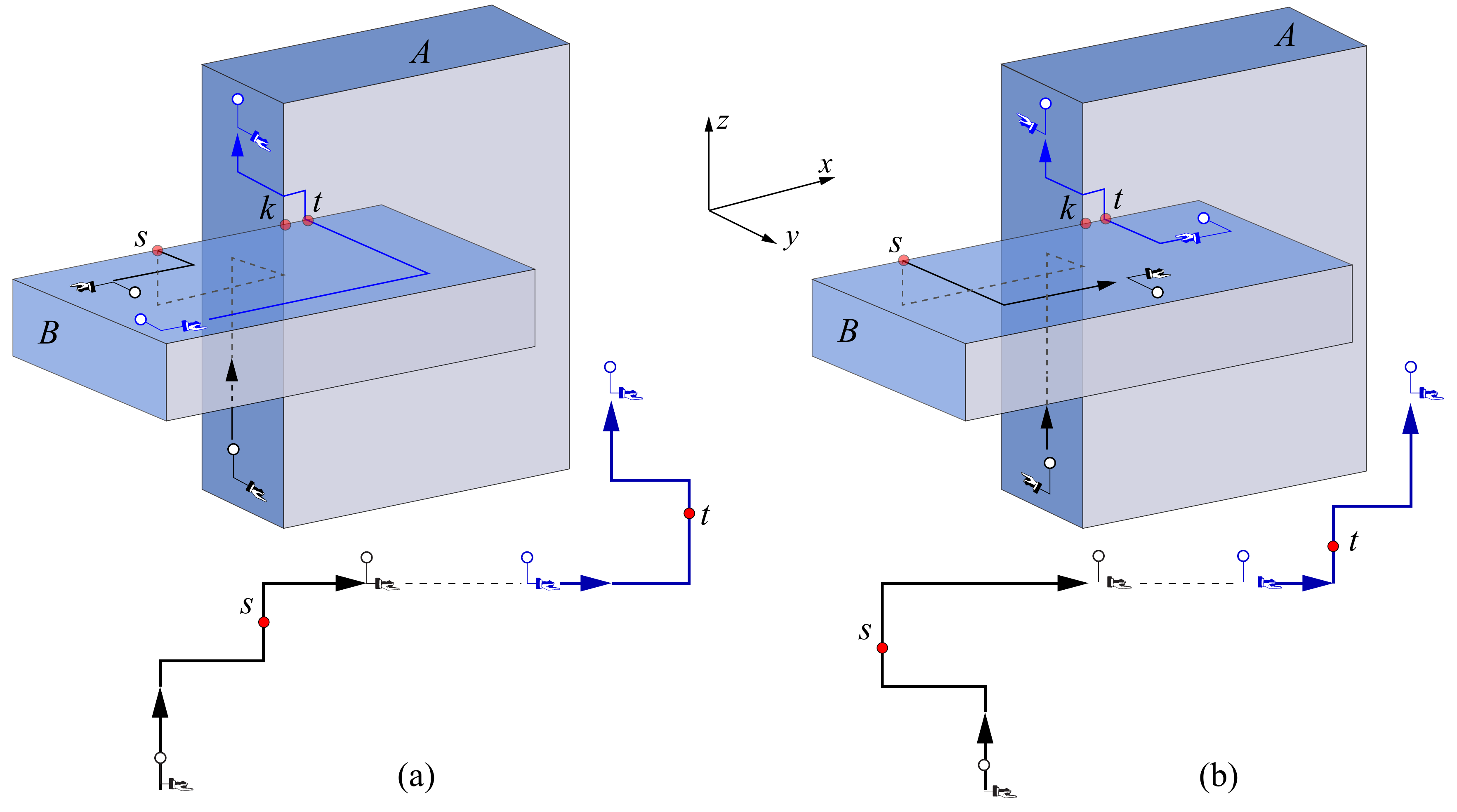}
\caption{C2-Transition paths; (a) \hand\ pointing toward $k$ (b) \hand\ pointing away from $k$.}
\label{fig:C2-transitions}
\end{figure}
%%%%%%%%%%%%%%%%%%%%%%%%%%%%%%%%%Figure End
%
\begin{enumerate}
\squeezelist
\item If the \hand\ on $A$ points toward $k$, then the forward transition from $A$ to $B$ uses a small L-shaped strip from the front/back face of $B$ containing $k$, as depicted in Fig.~\ref{fig:C2-transitions}a. (Follow the path from $A$ to $B$ in the direction of the arrows, ignoring for the moment the return path from $B$ to $A$.) Note that Fig.~\ref{fig:C2-transitions} depicts only the case where $B$ is a front child %attached to a left edge of $A$, 
and the unfolding proceeds clockwise along $A$. However, we will see that all other cases can be derived from this case.
\item If the \hand\ on $A$ points away from $k$, then the forward transition from $A$ to $B$ is similar to the transition from the previous case, with the unfolding going in opposite direction on $B$. The transition for this case is depicted in Fig.~\ref{fig:C2-transitions}b.
\end{enumerate}
Each of these forward transitions positions the \hand\ parallel to the $x$-axis on a horizontal face of $B$, so  invariant (I1) is satisfied. By Property~\ref{prop:handexit}, the \hand\ will be parallel to the $x$-axis after an $x$-unfolding of $B$, pointing in the direction opposite to the one upon entering $B$. This property enables the return transitions from $B$ to $A$ depicted in Fig.~\ref{fig:C2-transitions}.

%%%%%%%%%%%%%%%%%%%%%%%%%%%%%%%%%Figure Begin
\begin{figure}[htbp]
\centering
\includegraphics[width=0.9\linewidth]{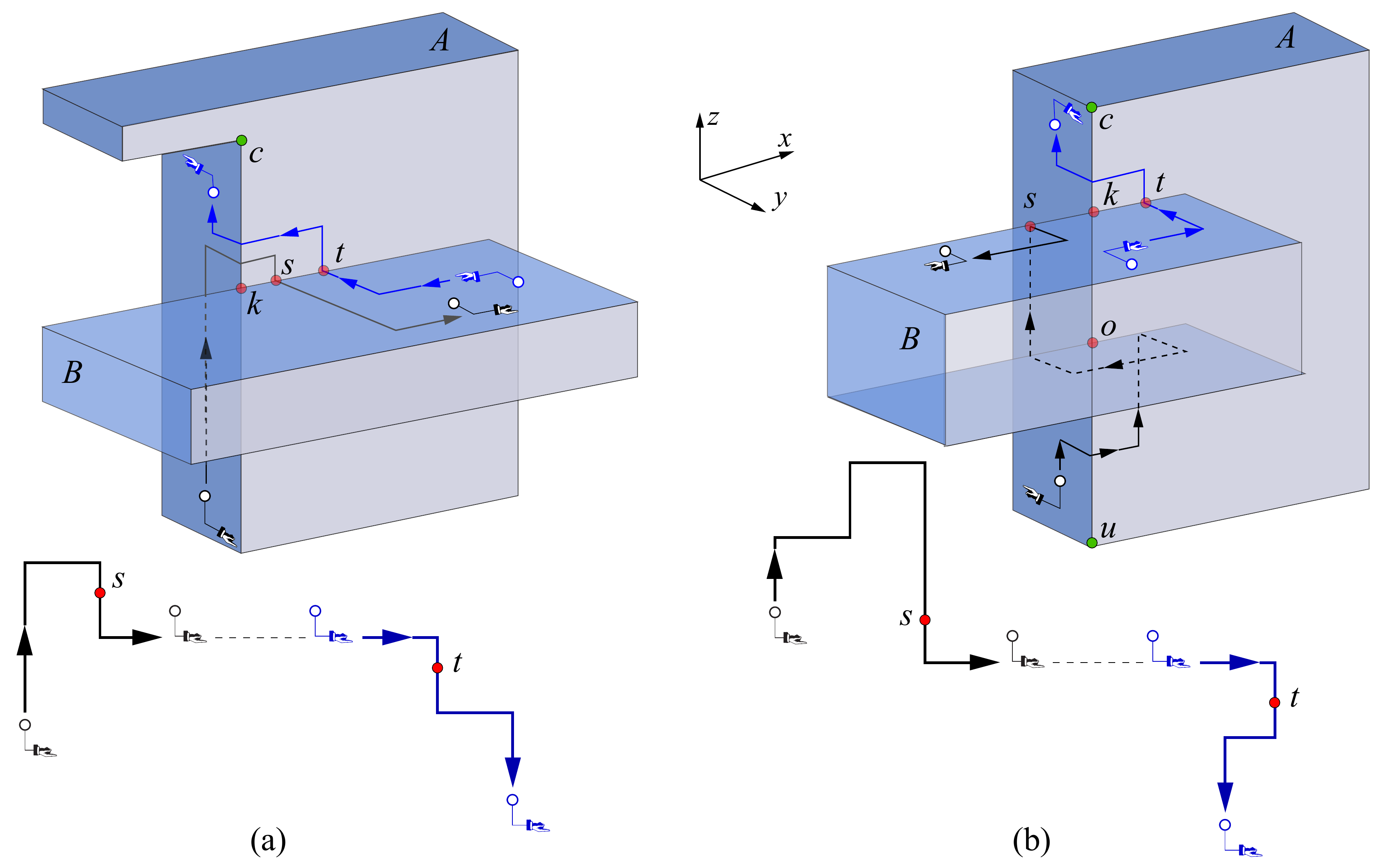}
\caption{Forward and return C2-transitions that reverse the orientation of the \hand\ on the parent band $A$ (a) \hand\ pointing toward $k$ (b) \hand\ pointing away from $k$, rim point $o$ exists.}
\label{fig:Reverse-C2-transitions}
\end{figure}
%%%%%%%%%%%%%%%%%%%%%%%%%%%%%%%%%Figure End

We will see that the unfolding algorithm (discussed later in \S\ref{sec:alg}) may sometimes require a C2-child band to flip the orientation of the \hand\ for the parent band, so that it is properly positioned for an upcoming forced turn (as required by invariant I3). In this case, both forward and return transitions are slightly different. Let $c$ be the corner vertex of $A$ such that $kc$ is on the rim of $A$ and extends from $k$ in the unfolding direction of $A$. By invariant (I3), a \hand\ reversal may be necessary in two situations: (i) if the \hand\ points toward $c$ and $c$ is reflex, and (ii) if the \hand\ points away from $c$, and $c$ is convex. The forward and return transitions for case (i) are depicted in Fig.~\ref{fig:Reverse-C2-transitions}a ($\xi_{3d}$ at the top and $\xi_{2d}$ at the bottom), and enable an $x$-unfolding of $B$. In case (ii) we use in the transition to $B$ a small neighborhood around a second point $o$ at the intersection between $A$ and $B$, as depicted in Fig.~\ref{fig:Reverse-C2-transitions}b. The intersection point $o$ always exists, because all left vertices are exposed (by the definition of an orthogrid). 
Note that the return transitions from Fig.~\ref{fig:Reverse-C2-transitions} reverse the orientation of the \hand\ on $A$, as required.

Figs.~\ref{fig:C2-transitions} and~\ref{fig:Reverse-C2-transitions} depict the situation in which the unfolding proceeded clockwise on $A$ upon reaching $k$.
The case when the unfolding strip proceeds counterclockwise on $A$ upon meeting $k$ is similar to the clockwise case, with forward and return path switching roles and traversed by the \hand\ in opposite direction. 
% (compare  Fig.~\ref{fig:mirror-C2-transitions}b from the appendix with Fig.~\ref{fig:C2-transitions}a).
%
Finally, a vertical $xz$-mirror of the above cases depicts the situation in which $B$ is a back child of $A$.

We are now ready to assign an unfolding label to $B$ consistent with the transitions identified in this section. Recall that $B$'s label depends on the unfolding direction and the orientation of the \hand\ on $A$. As in the case of the C1 anchor class, here we discuss the situation where  $B$ is a front child of $A$ and the front strip of $A$ runs clockwise just before reaching the anchor $k$ attaching $B$ to $A$; all other cases are symmetric. If the \hand\ points towards $k$ as it reaches $k$ (as depicted in Fig.~\ref{fig:C2-transitions}a), we assign to $B$ the label $\cc'_{s}$ to indicate the fact that $B$'s unfolding starts at the back of $B$ in counterclockwise direction, and the entering strip runs alongside the back rim of $B$. Note that this label is consistent with the transitions from Fig.~\ref{fig:C2-transitions}a. If the \hand\ points away from $k$ as it reaches $k$ (as depicted in Fig.~\ref{fig:C2-transitions}b), we assign to $B$ the label $\c'_{t}$ to indicate the fact that $B$'s unfolding starts at the back of $B$ in clockwise direction, and the exiting strip runs alongside the back rim of $B$.  This label is consistent with  the transitions from Fig.~\ref{fig:C2-transitions}b.

\begin{table}[htpb]
\begin{small}
\centering{
\begin{tabular}{|c|c|c|c|c|}
\hline
Anchor Class & Child $B$ & Unfolding label for $A$ &  \hand\ orientation on $A$ & Unfolding label for $B$ \\
\hline\hline
\multirow{8}{*}{C2} &
\multirow{4}{*}{front} &
$\c_{s}$, $\c'_{t}$, $\cc'_{s}$, or $\cc_{t}$ &
Towards anchor & $\cc'_{s}$ (Figs.~\ref{fig:C2-transitions}a) \\
\cline{4-5}
& & (front strip clockwise)  & Away from anchor  & $\c'_t$ (Fig.~\ref{fig:C2-transitions}b)\\
\cline{3-5}
&& $\c'_{s}$, $\c_{t}$, $\cc_{s}$, or $\cc'_{t}$ & 
Towards anchor & $\cc'_t$ \\
\cline{4-5}
&&(front strip counterclockwise) & Away from anchor & $\c'_s$\\
\cline{2-5}
&\multirow{4}{*}{back} &
$\c'_{s}$, $\c_{t}$, $\cc_{s}$, or $\cc'_{t}$ & 
Towards anchor & $\cc_s$ \\
\cline{4-5}
&&(back strip clockwise) & Away from anchor & $\c_t$ \\
\cline{3-5}
&& $\c_{s}$, $\c'_{t}$, $\cc'_{s}$, or $\cc_{t}$ & 
Towards anchor &  $\cc_t$ \\
\cline{4-5}
&&(back strip counterclockwise) & Away from anchor & $\c_s$ \\
\hline
\end{tabular}
\caption{Unfolding labels for $B$ when attached to $A$ by a C2-anchor. }
\label{tab:C2labels}
}%centering
\end{small}
\end{table}

\noindent
The unfolding labels for the situation when $B$ is a front child of $A$ and the front strip of $A$ runs counterclockwise just before reaching $k$  are derived from the unfolding labels for the situation discussed above, with $s$ and $t$ switching roles (because the forward and return transition strips switch roles). Finally, the cases where $B$ is a back child of $A$ are $xz$-mirrors of the corresponding cases with $B$ as a front child, and the unfolding labels can be obtained by simply dropping the prime symbol to indicate that $B$'s unfolding starts on the from rim of $B$. The unfolding labels for all cases are listed in Table~\ref{tab:C2labels}.

\section{Unfolding a Child Band}
\label{sec:unfchild}
\medskip
\noindent
Later in \S\ref{sec:alg} we will define our main unfolding procedure \unfoldrims\ for a band $A$. This procedure unfolds $A$ and all its descendants in the unfolding tree $T$, starting at entering point $s$ on $A$ and ending at exiting point $t$ on $A$. We assume that the \hand\ is already positioned at entering point $s$ on $A$ at the beginning of the \unfoldrims\  procedure, and the clockwise/counterclockwise direction of $A$ is determined by $A$'s unfolding label. After recursively unfolding all descendants of $A$ in $T$ (details to be filled in later), the procedure ends at $t$ with the \hand\ pointing in the direction opposite to the one indicated by the entering \hand. 
%
%parallel to the $x$-axis (in the case of \rxunfold) or to the $y$-axis (in the case of \ryunfold). By Property~\ref{prop:handexit}, the exiting hand points in the direction opposite to the one indicated by the entering hand.

In this section we make use of this \unfoldrims\ procedure and the tools developed in the previous section to recursively unfold a child $B$ of a parent band $A$, as described in the algorithm \unfold\ from Table~\ref{tab:C1unfold}. % (note that it uses the  \ryunfold\ procedure briefly discussed above).
The case when $B$ is a C2-child that must reverse the direction of the \hand\ on $A$ is handled by the procedure \unfoldrev\ defined in Table~\ref{tab:C2unfoldrev}.

\begin{table}[hptb]
\begin{center}
\vspace{1mm} \fbox{
\begin{minipage}[h]{0.85\linewidth}
\centerline{\unfold(Parent Band $A$, Child Band $B$)}
\vspace{1mm}{\hrule width\linewidth}\vspace{2mm} %\vbox{\hrule width\linewidth}
\small{\begin{tabbing}
........\=......\=.......\=.......\=..................................................\kill
Let $k$ be the anchor point connecting $B$ to the rim $r$ of $A$ \\
If $B$ is a C1-child of $A$ \\
\> Let $c$ be the corner vertex of $B$ such that $kc$ is vertical and crosses $r[A]$ \\
\> Execute forward C1-transition (Fig.~\ref{fig:C1-transitions}a for $c$ convex, Fig.~\ref{fig:C1-transitions}b for $c$ reflex)\\
\> \unfoldrims($B$) \\
\> Execute return C1-transition (Fig.~\ref{fig:C1-transitions}a for $c$ convex, Fig.~\ref{fig:C1-transitions}b for $c$ reflex) \\
If $B$ is a C2-child of $A$ \\
\> Execute forward C2-transition as in Fig.~\ref{fig:C2-transitions}a (\ref{fig:C2-transitions}b) if \hand\ points to (away from) $k$\\
\> \unfoldrims($B$) \\
\> Execute return C2-transition as in Fig.~\ref{fig:C2-transitions}a (\ref{fig:C2-transitions}b) if \hand\ points to (away from) $k$ \\
Mark $B$ visited
\end{tabbing}}
\end{minipage}
}%fbox
\vspace{1mm}
\end{center}
\vspace{-1em}\caption{Recursive unfolding of a child band.}
\label{tab:C1unfold}
\end{table}

\begin{table}[hptb]
\begin{center}
\vspace{1mm} \fbox{
\begin{minipage}[h]{0.85\linewidth}
\centerline{\unfoldrev(Parent Band $A$, C2-child Band $B$)}
\vspace{1mm}{\hrule width\linewidth}\vspace{2mm} %\vbox{\hrule width\linewidth}
\small{\begin{tabbing}
......\=......\=.......\=.......\=..................................................\kill
Let $k$ be the anchor point connecting $B$ to $A$. \\
Execute forward C2-transition as in Fig.~\ref{fig:Reverse-C2-transitions}a (\ref{fig:Reverse-C2-transitions}b) if \hand\ points to (away from) $k$\\
\unfoldrims($B$) \\
Execute return C2-transition as in Fig.~\ref{fig:Reverse-C2-transitions}a (\ref{fig:Reverse-C2-transitions}b) if \hand\ points to (away from) $k$ \\
Mark $B$ visited
\end{tabbing}}
\end{minipage}
}%fbox
\vspace{1mm}
\end{center}
\vspace{-1em}\caption{Unfolding a C2-child that reverses the \hand\ orientation on the parent.}
\label{tab:C2unfoldrev}
\end{table}

\section{Unfolding Rim Children}
Having developed the procedure to recursively unfold a single child $B$ of a parent band $A$, we now proceed to unfolding all children anchored to a rim of a band $A$, as described in the \unfoldrim\ algorithm from Table~\ref{tab:rimunfold}. 
%Here we emphasize the ``unvisited'' term, because some children will be marked visited and therefore will be ignored by this procedure. 
This procedure will be invoked for both the entering and the turnaround rim of $A$. In the case of the turnaround rim $r$ of $A$, the \unfoldrim\ procedure may need to handle one special child marked \emph{turnaround} by the main unfolding procedure. This happens when $A$ has children anchored to $r$, so a vertical strip from $r[A]$ may be unavailable for the turnaround (as in the single box unfolding discussed in \S\ref{sec:kinunf}). In this case one particular child $B$ of $A$ attached to $r$ will be charged with reversing the unfolding direction of $A$. (Note that the unfolding strip resulting from the recursive unfolding of $B$ plays the role of the turnaround strip in the single band unfolding from Fig.~\ref{fig:Unf0}b.)
Let \unfoldret\ be a procedure identical to \unfold, but with each return transition modified so that it retraces the forward transition. The result is that the \hand\ reverses its direction on the parent band. This procedure will be used in unfolding turnaround children. 

At the start of the \unfoldrim\ procedure, the \hand\ can be on a horizontal face of $A$ parallel to the $x$-axis (by invariant I1), or on a vertical face of $A$ parallel to the $y$-axis or the $z$-axis (by invariant I2).  
%We say that the \hand\ \emph{reaches} an anchor $\sigma$ when $\sigma$ is incident to a point $u$ on the front rim, and the \hand\ is about to pass alongside $u$. 
The \hand\ begins cycling around $r$, and processes three types of events that occur during this cycle. The actions taken in response to each of these events are summarized in Table~\ref{tab:rimunfold}. Next we discuss the details of each event.

\begin{table}[hptb]
\begin{center}
\vspace{1mm} \fbox{
\begin{minipage}[h]{0.85\linewidth}
\centerline{\unfoldrim(Band A, Rim r)}
\vspace{1mm}{\hrule width\linewidth}\vspace{2mm} %\vbox{\hrule width\linewidth}
\small{\begin{tabbing}
....\=......\=.........\=....................\=.......\=..................................................\kill
Proceed with the \hand\ sliding along $r$ in the direction indicated by $A$'s unfolding label. \\
While cycling around $r$, process the following events: \\
\>Let $I$ be the band face of $A$ that the \hand\ is currently on. \\
\>Let $e_I$ be the edge of $I$ on the rim $r$. \\
\>Let $c$ be the corner vertex of $r$ that will be next encountered by the \hand.\\
\>Let $J$ ($e_J$) be the other face of $A$ (edge of $J$) incident to $c$ \\
\\
\>If the hand indicator is about to pass the corner vertex $c$ /* Event 1 */ \\
\>\> If ~($I$ is horizontal and there are unvisited children of $A$ attached to $e_J$) OR \\
\>\> ~~~~($I$ is vertical and the \hand\ is parallel to the $y$-axis) \\
\>\>\> Execute a forced turn at $c$. \\
\\
\>If the \hand\ reaches an anchor $k$ attaching an unvisited child $B$ of $A$ to $A$ /* Event 2 */\\
\> \>If $B$ is marked \emph{turnaround}, then \unfoldret($A$, $B$) \\
\>\> Else If $B$ is a C1-child of $A$ then \unfold($A$, $B$) \\
\>\> Else If $B$ is a C2-child of $A$ ~~~/* $I$ is vertical, $J$ is horizontal */ \\
\>\>\> If ~(no unvisited child of $A$ is anchored on $kc$) AND \\
\>\>\> ~~~ (Invariant I3 is violated by the \hand\ orientation relative to $c$) \\
\>\>\>\> \unfoldrev(A, B) \\
\>\>\> Else \unfold(A, B). \\
\\
\>If there are no more unvisited children of $A$ attached to $r$ /* Event 3 */\\
\>\> If $I$ is horizontal, then exit. \\
\>\> If $r$ is the first rim of $A$ visited OR $r$ is the turnaround rim of $A$  \\
\>\> \> If the \hand\ is parallel to the $y$-axis, proceed to $c$ then force a turn at $c$. \\
\>\> \> Else proceed in the direction of the \hand\ to $J$, then exit.
\end{tabbing}}
\end{minipage}
}%fbox
\vspace{1mm}
\end{center}
\vspace{-1em}\caption{Unfolding all unvisited children attached to a band rim.}
\label{tab:rimunfold}
\end{table}

\paragraph{Event 1.} The first type of event occurs when the \hand\ reaches a corner vertex $c$ of $A$. Let $I$ and $J$ be the two faces of $A$ incident to $c$, with the \hand\ currently on $I$. At this point, the \hand\ is about to make a transition from $I$ to $J$. 
If $I$ is horizontal, then the \hand\ is parallel to the $x$-axis. If the vertical rim edge $e_J$ incident to $c$ contains no anchor points attaching an unvisited child of $A$, then no action is taken: $A$'s unfolding proceeds onto $J$, with the \hand\ parallel to the $z$-axis.
Otherwise, if $e_J$ contains at least one anchor point attaching an unvisited child of $A$, then the \hand\ must be reoriented parallel to the $y$-axis on $J$, so that the invariant (I2) holds. This is accomplished by a forced turn at $c$.

If $I$ is vertical, invariant (I2) tells us that the \hand\ may be parallel to the $z$-axis or to the $y$-axis. In the first case, no action is necessary: the \hand\ continues to $J$, where it becomes parallel to the $x$-axis, so invariant (I1) is maintained.
In the latter case, one of the forced turns depicted in Figs.~\ref{fig:turns}b and~\ref{fig:turns2}b must be taken at $c$, depending on whether $c$ is convex or reflex. By invariant (I3), such a forced turn at $c$ reorients the \hand\ parallel to the $x$-axis, pointing in the unfolding direction. Thus invariant (I1) is maintained.

\begin{figure}[htbp]
\centering
\includegraphics[width=0.8\linewidth]{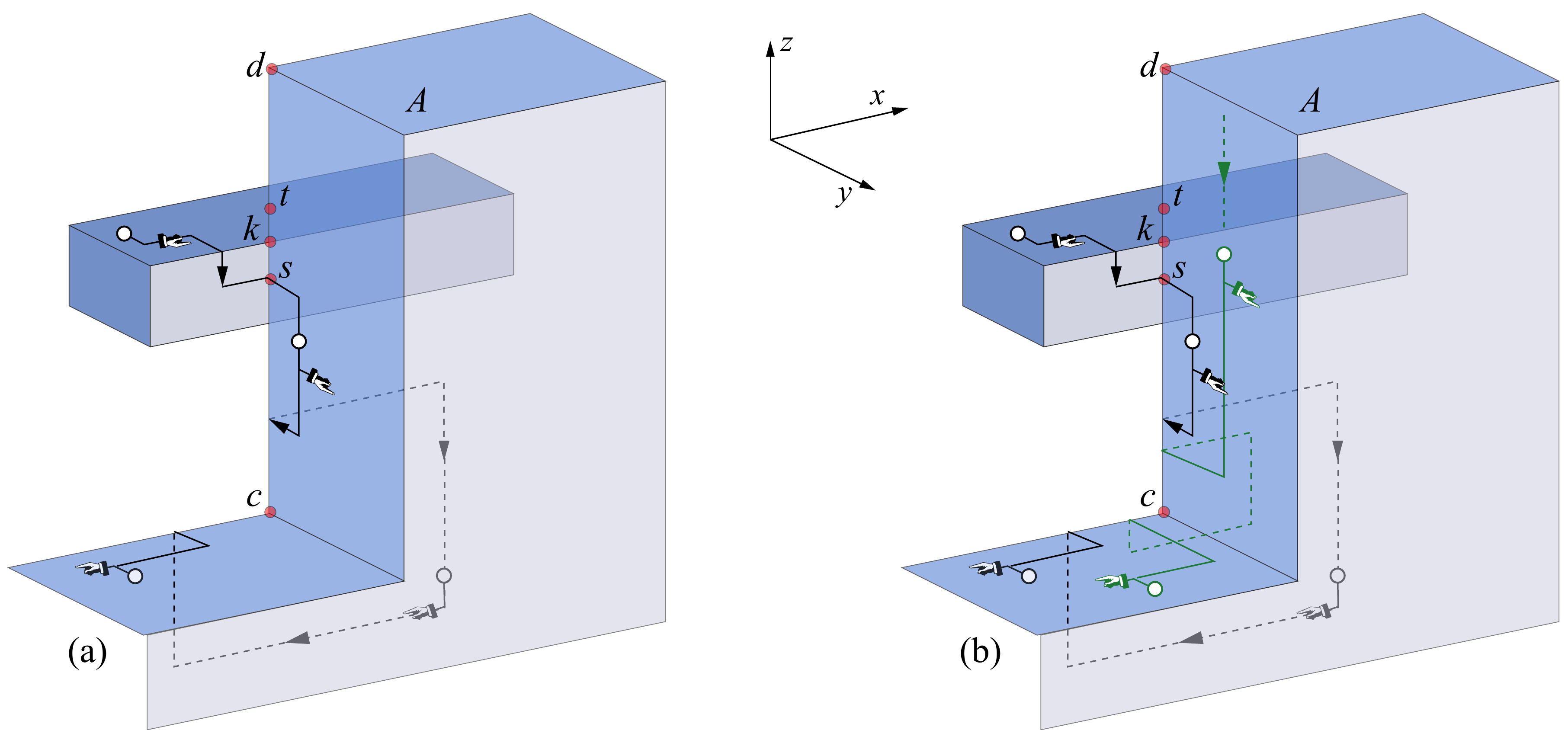}
\caption{Transition to $A$ on a vertical rim edge (a) First visit of $c$ (b) Second visit of $c$.}
\label{fig:verticalstart}
\end{figure}

If this is the first time that $c$ gets encountered by the \hand, then the unfolding strip marks the first forced turn at $c$. It is however possible that this is the second time the \hand\ encounters $c$ while positioned parallel to the $y$-axis; this situation occurs when the unfolding of $A$ starts at a point $s$ on $e_I$ and there are children attached to $e_I$ on both sides of $s$ (see Fig.~\ref{fig:verticalstart}). In this case the \hand\ simply traces the path taken by the first forced turn at $c$, so that it becomes parallel to the $x$-axis on $J$. 

%%%%%%%%%%%%%%%%%%%%%%%%%%%%%%%%%Figure Begin

In the example from Fig.~\ref{fig:verticalstart}, the anchor point  $k$ that attaches $A$ to its parent is located on the vertical edge $cd$ of $A$. The unfolding of $A$ begins at point $s$ located on the vertical segment $kc$ just below $k$, and proceeds downwards with the \hand\ parallel to the $y$-axis; children attached to $A$ by anchor points located on the rim segment $sc$ are recursively unfolded as they are encountered on the way down to $c$. A forced turn is taken at corner vertex $c$ to reorient the \hand\ parallel to the $x$-axis. Now note that there may be other children of $A$ attached to $A$ by anchor points located on the rim segment $kd$ above $k$. These children get visited as the unfolding strip moves downwards from $d$, after a forced turn at $d$ that reorients the \hand\ parallel to the $y$-axis. Once the unfolding strip passes $s$ on the way down to $c$, it simply retraces the path previously taken around $c$, until it reaches the the next horizontal face of $A$. At this point, the unfolding strip has visited all children of $A$ and the 
\unfoldrim\ procedure terminates. 

\paragraph{Event 2.} The second type of event occurs when the \hand\ reaches an anchor attaching an unvisited child $B$ of $A$. At this point the parent band unfolding is suspended and $B$ is recursively unfolded. If $B$ is a turnaround child, the unfolding direction of $A$ is reversed; otherwise the unfolding resumes on $A$ from the point it left off. If $B$ is a C2-child, then the decision whether $B$ must reverse the orientation of the \hand\ on $A$ so as to satisfy invariant (I3) must be taken at this point. 

% One decision made at this point is whether a C2-child $B$ must reverse the orientation of the \hand\ on $A$ or not, so that invariant (I3) is satisfied.

\paragraph{Event 3.} The third and final type of event occurs when all rim children of $A$ have been visited. In this case it may be necessary to proceed to the closest horizontal face of $A$, to ensure that the \hand\ ends up parallel to the $x$-axis. This is necessary if the unfolding strip needs to follow the unfolding path towards the other rim of $A$ (as in the single box unfolding from S\ref{sec:kinunf}) in order to complete the unfolding of $A$.

% what if you step past the child that needs to reverse the unfolding direction for turnaround

\section{Main Unfolding Algorithm}
\label{sec:alg}
In this section we describe the algorithm for recursively unfolding a band and all its descendants. Let $A$ be a band to unfold, initially the root band. The unfolding of $A$ begins at entering point $s$ and follows the basic unfolding path tied to $A$'s unfolding label, as discussed in \S\ref{sec:kinunf}. Let $r$ be the rim of $A$ containing $s$, and let $r'$ be the opposite (turnaround) rim of $A$. The procedure \unfoldrims\ from Table~\ref{tab:unfoldrims} describes our method for recursively unfolding all children attached to the band $A$, starting at the entering point $s$ on $A$. In the following we briefly discuss this procedure.

\begin{table}[p]
\begin{center}
\vspace{1mm} \fbox{
\begin{minipage}[h]{0.9\linewidth}
\centerline{\unfoldrims(Band $A$)}
\vspace{1mm}{\hrule width\linewidth}\vspace{2mm} %\vbox{\hrule width\linewidth}
\small{\begin{tabbing}
.......\=......\=......\=.......\=.......\=..................................................\kill
 \> Let $r$ be the entering rim and $r'$ the turnaround rim of $A$. \\
\> \emph{If $A$'s unfolding label if of type $*_s$:} \\
1 \>  \> Proceed with the \hand\ pointing in the direction indicated by $A$'s unfolding label. \\
2 \>   \> \unfoldrim($A$, $r$) \\
3 \>   \> Move in the $y$-direction toward $r'$ (see box unfolding, \S\ref{sec:boxunf}). \\
4 \>   \> If no children of $A$ are attached to $r'$ \\
   \> \> \> Turn around by moving vertically across the facet $r'[A]$ (see box unfolding, \S\ref{sec:boxunf}). \\
  \> \> Else \\
   \> \> \> Proceed in the direction of the \hand\ up to the first child $B$ of $A$ anchored on $r'$. \\
   \> \> \> (If $B$ is a C2-child, a forced turn is taken at the last vertex encountered before $B$). \\
 \>   \> \> Reassign $B$'s unfolding subscript to $*_s$, so that it matches $A$'s unfolding subscript. \\
 \>   \> \> \unfoldret($A$, $B$) /* turnaround executed here */ \\
 \>   \> \> \unfoldrim($A$, $r'$) \\
 \>   \> Retrace unfolding path on $A$ back to $r$ \\
   \>\emph{If $A$'s unfolding label is of type $*_t$:} \\
5 \>   \> From $s$, move parallel to the $y$-axis toward $r'$, then make a full cycle around $A$. \\
6\>   \>  Move in the $y$-direction toward $r'$ (see box unfolding, \S\ref{sec:boxunf}). \\
7 \>   \> If no children of $A$ are attached to $r'$ \\
  \> \> \> Turn around by moving vertically across the facet $r'[A]$ (see box unfolding, \S\ref{sec:boxunf}). \\
  \>   \> Else \\
  \> \> \> Let $B$ be the child of $A$ anchored on $r'$ that would be last encountered in a cycle \\
  \> \> \> \> around $A$, starting from the current \hand\ position. \\
  \> \> \> Mark $B$ \emph{turnaround}. Reassign $B$'s unfolding subscript to $*_t$ to match $A$'s subscript. \\
  \> \> \> \unfoldrim($A$, $r'$) \\
8 \>   \> Retrace unfolding path on $A$ back to $r$. \\
9 \>   \> \unfoldrim($A$, $r$) \\
10 \> \emph{If $A$ is a child of type} C1 and the \hand\ is parallel to the $x$-axis \\
\> \> /* must redirect exiting hand parallel to the $y$-axis */ \\
   \> \> Continue cycling up to the last corner $c$ of $A$ encountered before exit point $t$ \\
   \> \> Retrace the forced turn at $c$ (see unfolding variations, \S\ref{sec:kinunf})\\
   \> \> Retrace the entering path back to exit point $t$ \\
 \> \emph{Else} \\
   \> \> Continue cycling in the direction of the \hand\ up to exit point $t$ 
\end{tabbing}}
\end{minipage}
}%fbox
\vspace{1mm}
\end{center}
\vspace{-1em}\caption{Main unfolding algorithm for a band and all its descendants.}
\label{tab:unfoldrims}
\end{table}

If $A$ has an $*_s$ label, then $A$'s entering spiral visits $A$'s children attached to $r$, and $A$'s exiting spiral visits $A$'s children attached to $r'$. After recursively unfolding all $A$'s children anchored on $r$ (step 2), the spiral proceeds towards the back rim $r'$ (step 3) where it needs to turn around and reverse the unfolding direction for $A$. If $A$ has no children attached to $r'$, then the facet $r[A]$ is exposed\footnote{This statement is established by Lem.~3 in~\cite{Damian-Flatland-O'Rourke-2007-epsilon}, with anchors here playing the role of beams in~\cite{Damian-Flatland-O'Rourke-2007-epsilon}.}, therefore a vertical strip from $r[A]$ can be used for the turnaround, as in the single band unfolding case (see the if branch of step 4). Note however that such a vertical strip may be unavailable if $B$ has children attached to $r'$. We resolve this issue by using a child $B$ attached to $r'$ to reverse the unfolding direction of $A$. Thus the entering spiral must continue cycling in the unfolding direction up to the first child encountered, which reverses $A$'s unfolding direction as required (see the else branch of step 4). Note that $B$'s unfolding subscript must be identical the $A$'s unfolding subscript, so that the relative position of entering and exiting spiral is the same in both (it may help to view the entering spiral of $B$ as an extension of the entering spiral of $A$, and the exiting spiral of $A$ as an extension of the exiting spiral of $B$). After $B$ is recursively unfolded, the exiting spiral for $A$ moves alongside the turnaround rim $r'$ (because $A$'s label is $*_s$ ) and recursively unfolds all children attached to $r'$,  then retraces the path on $A$ back to $r$.

If $A$ has a $*_t$ label, then $A$'s exiting spiral visits $A$'s children attached to $r$, and $A$'s entering spiral visits $A$'s children attached to $r'$. In this case, the child $B$ of $A$ \emph{last} visited by the entering spiral reverses the unfolding direction for $A$, after all other children attached to $r'$ have been recursively unfolded by the entering spiral (see the else branch of step 7). This way, the exiting spiral can proceed immediately back to $r$ after the turnaround (step 8), and continue unfolding all children attached to $r$ (step 9). 

The last strip of the exiting spiral depends on whether the exit point $t$ is located on a horizontal or a vertical rim edge of $A$. If $t$ is located on a horizontal rim edge, then the \hand\ simply cycles around in the direction of the \hand\ back to $t$ (see the else branch of step 10). If $t$ is on a vertical rim edge of $A$, it must be that $A$ is a C1-child and the exiting hand must be repositioned parallel to the $y$-axis, so that Property~\ref{prop:handexit} is satisfied.  (One such situation is depicted in Fig.~\ref{fig:verticalstart}.) This can be accomplished by retracing the very first forced turn taken on $A$, as described in \S\ref{sec:kinunf} (see the if branch of step 10). 
If $root$ is the root band of the unfolding tree $T$ corresponding to an orthogrid $\P$, then \unfoldrims($root$) produces an unfolding spiral $\xi_{3d}$ cycling around all band faces of $\P$. 

\section{Attaching Front and Back Faces}
\label{sec:frontback}
The planar piece $\xi_{2d}$ has the basic staircase shape illustrated in Fig.~\ref{fig:Unf0}, with only two exceptions.
The first exception occurs at the forced turns around corner vertices depicted in Fig.~\ref{fig:turns2}. We 
thicken the spiral pieces corresponding to forced turns so that no front/back surface piece gets trapped 
between such a spiral piece and the band faces incident to the turn  vertex. 
The second exception occurs at the transitions between bands depicted in 
Figs~\ref{fig:C1-transitions}a,~\ref{fig:C2-transitions} and~\ref{fig:Reverse-C2-transitions}b. Again we 
thicken the spiral pieces corresponding to transitions between bands so that no front/back surface piece 
gets trapped between such a spiral piece and the band faces involved in the transitions. 
The remaining  pieces of $\xi_{3d}$ (which correspond to monotone pieces of $\xi_{2d}$) can be thickened in the $y$-direction without overlap so that $\xi_{3d}$ covers all of the top, bottom, left and right faces of the orthogrid.
%It also covers the strips from the front/back faces used in forced turns (Fig.~\ref{fig:turns2}), transitions between bands (Figs~\ref{fig:C1-transitions},~\ref{fig:C2-transitions} and~\ref{fig:Reverse-C2-transitions}), and turnaround to reverse the spiral direction in the base box unfolding (Fig.~\ref{fig:Unf0}). 

We now show that the remaining front and back surface pieces that are not part of $\xi_{3d}$ 
can be attached orthogonally to $\xi_{2d}$ without overlap.
Consider the horizontal boundary segments of $\xi_{2d}$ that correspond to boundary segments of $\xi_{3d}$ adjacent to front and back surface pieces that are not part of $\xi_{3d}$.  Imagine these boundary segments emanate lightrays downwards. Then all unattached front/back pieces are illuminated. It can be verified that these pieces can now be attached to the corresponding illuminating rim segments of $\xi_{2d}$ without overlap. 

\section{Future Work}
We believe that the techniques described in this paper can be extended to unfold any orthogonal polyhedra of genus zero with constant refinement of the gridded surface. However, the details involved in handling unexposed left vertices are much more involved. 

%\bibliographystyle{alpha}
%\bibliography{unfolding}

\newcommand{\etalchar}[1]{$^{#1}$}

\end{document}